\documentclass[twocolumn,a4paper]{article}
\usepackage{epsfig}

\newcommand{\be}{\begin{equation}}
\newcommand{\ee}{\end{equation}}
\newcommand{\ba}{\begin{eqnarray}}
\newcommand{\ea}{\end{eqnarray}}
\newcommand{\ban}{\begin{eqnarray*}}
\newcommand{\ean}{\end{eqnarray*}}

\newcommand{\ket}[1]{\mbox{$ | #1 \rangle $}}

\begin{document}

\title{\large\bf Imbedding nonlocality in a relativistic chronology}

\author{\small Antoine Suarez\thanks{suarez@leman.ch}\\{\it\small Center for Quantum
Philosophy, P.O. Box 304, CH-8044 Zurich, Switzerland}}

\date{\small August 9, 2001}

\maketitle

{\footnotesize \emph{Abstract}: An alternative description
imbedding nonlocality in a relativistic chronology is proposed. It
is argued that vindication of Quantum Mechanics in experiments
with moving beam-splitters would mean that there is no real time
ordering behind the nonlocal correlations}

 \vspace{2mm}

The relationship between Quantum Mechanics and Relativity has been
object of vast analysis since John Bell showed that: a) if one
only admits relativistic local causality (causal links with $v\leq
c$), the correlations occurring in two-particle experiments should
fulfill clear locality conditions (``Bell's inequalities''), and
b) for these experiments Quantum Mechanics bears predictions
violating such locality criteria (``Bell's theorem'') \cite{jb64}.
Bell-type experiments conducted in the past two decades, in spite
of their loopholes, suggest a violation of local causality:
statistical correlations are found in space-like separated
detections; violation of Bell's inequalities ensure that these
correlations are not pre-determined by local events. Nature seems
to behave nonlocally, and Quantum Mechanics predicts well the
observed distributions. Nevertheless, nonlocality (``Bell
influences'') cannot be used for faster-than-light communication.

Nonlocal correlations cannot just appear by chance: they require
an ordering of the events, causality in some sense. But we use to
think about causality as related to some temporal sequence.
Therefore, also taking nonlocality for granted, the important
question remains: is there a time ordering behind the nonlocal
correlations?

Bohm's theory proposes to imbed Quantum Mechanics in a preferred
frame or absolute time, in which one event is caused by some
earlier event \cite{dbbh}. The theory does not make predictions
conflicting with Quantum Mechanics but is rather a particular
interpretation of it. However, it does not tell us how to trace
this frame \cite{jb64}, so that one sees no mean to decide whether
the bohmian ``quantum ether'' has any physical reality at all.

Recent work shows that it is possible to imbed nonlocality in a
real relativistic time ordering, providing one gives up Quantum
Mechanics in a new type of experiments involving moving devices.
This happens within Multisimultaneity, a nonlocal description
using many frames to establish the cause-effect links
\cite{asvs97.1, as00.1}. More specifically these frames are
supposed to be those of the beam-splitters (``choice-devices'')
\cite{ophoc}. Within each frame the links always correspond to a
well defined chronology, one event never depending on some future
event. Multisimultaneity has already been developed in the context
of 2-particle experiments with moving beam-splitters. In this
article we implement it in 3-particle ones, and discuss the
meaning of a possible vindication of Quantum Mechanics in tests
using moving beam-splitters.

\begin{figure}[t]
\centering\epsfig{figure=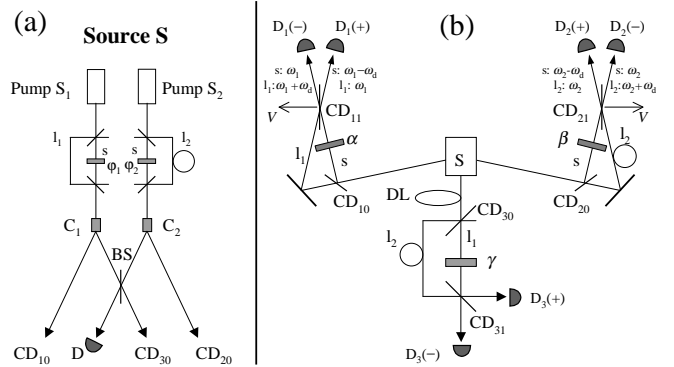,width=86mm}
{\caption{\footnotesize (a) Diagram of the source S used in the
experiment. (b) 3-particle experiment using moving choice-devices
CD$_{11}$ and CD$_{21}$. See text for detailed description.}}
\label{Seriesfig1}
\end{figure}

In Fig. 1 (a) is sketched the schema of a source S capable of
producing maximally energy-time entangled photon triplets
\cite{bgtz00}. Photons coming from the pulsed pump laser S$_{1}$
reach the nonlinear crystal C$_{1}$ either by path $l_{1}$ or path
$s$, and those from the pulsed pump S$_{2}$ reach the nonlinear
crystal C$_{2}$ either by $l_{2}$ or $s$. At C$_{1}$ and C$_{2}$
twin-photons are created by parametric down-conversion. Two output
beams, one of C$_{1}$ and one of C$_{2}$, are directly guided to
the beam-splitters CD$_{10}$, respectively CD$_{20}$, and as
represented in Fig. 1 (b) illuminate two interferometers which use
moving choice-devices CD$_{11}$ and CD$_{21}$. The other two beams
are led to interfere into beam-splitter BS. One of the output
ports of BS is monitored by detector D, and the photons leaving by
the other are guided to beam-splitter CD$_{30}$ and, as
represented in Fig. 1 (b), illuminate a third interferometer using
a resting beam-splitter CD$_{31}$. The location of this device can
be adjusted by means of delay line DL.

All CD$_{il}$ ($i\in\{1,2,3\}, l\in\{0,1\}$) are assumed to be
50-50 beam-splitters. The two output ports of each CD$_{i1}$ are
monitored by detectors D$_{i}(\sigma)$ ($\sigma\in\{+,-\}$). The
short arms of the two interferometers within the source S, and the
short arms of the interferometers 1 and 2, are all of them
supposed to be equal in length. The arms $l_{1}$ and $l_{2}$ of
interferometer 3 are equal to the arms $l_{1}$ respectively
$l_{2}$ within the source S. The phase parameters are labeled
$\varphi_{1}$, $\varphi_{2}$, $\alpha$, $\beta$, and $\gamma$.

We consider only the cases in which detector D registers a photon
traveling by one of the short arms, and there is one photon in
each of the three output ports leading to CD$_{10}$, CD$_{20}$,
CD$_{30}$. We assume the pumps S$_{1}$ and S$_{2}$  to work well
synchronized so that the detected photon triplets signals in
 D$_{i}(\sigma)$ will exhibit the same time difference for the paths
($S_{1}sl_{1},S_{2}l_{2}s,S_{2}l_{2}l_{1}$) and
($S_{1}l_{1}s,S_{2}sl_{2},S_{1}l_{1}l_{2}$), where the first path
expression denotes that: photon 1 comes from pump S$_{1}$, travels
path $s$ within the source, and then $l_{1}$ in interferometer 1;
photon 2 comes from pump S$_{2}$, travels first path $s$ within
the source, and then $l_{2}$ in interferometer 2; photon 3 comes
from pump $S_{2}$, travels path $l_{2}$ within the source, and
then $l_{1}$ of interferometer 3; and the second path expression
has a similar meaning \cite{bgtz00}. Because of the movement of
CD$_{11}$ and CD$_{21}$ the frequency of the reflected photons is
Doppler-shifted by an amount $\omega_{d}$, but the setup in Fig. 1
(b) is arranged so that the total frequency shift for each of the
two paths is the same. Therefore detection of the triplets
traveling the paths ($S_{1}sl_{1},S_{2}l_{2}s,S_{2}l_{2}l_{1}$)
and ($S_{1}l_{1}s,S_{2}sl_{2},S_{1}l_{1}l_{2}$) will exhibit
interferences.

According to Quantum Mechanics, beyond the devices CD$_{10}$,
CD$_{20}$ and CD$_{30}$ the three particles are in a GHZ-state of
the form:

\begin{footnotesize}
\ba
\ket{\psi}=\frac{1}{\sqrt{2}}(\ket{S_{1}sl_{1}}\ket{S_{2}l_{2}s}\ket{S_{2}l_{2}l_{1}}
+e^\phi\ket{S_{1}l_{1}s}\ket{S_{2}sl_{2}}\ket{S_{1}l_{1}l_{2}})
\label{psi} \ea
\end{footnotesize}

\noindent where $\phi$ is a phase factor.

Independently of any timing it holds that:

\begin{footnotesize}
\ba &&Pr^{QM}(\rho,\sigma,\omega)
=\frac{1}{K}|A(S_{1}sl_{1}\rho,S_{2}l_{2}s\sigma,S_{2}l_{2}l_{1}\omega)\nonumber\\
&&\hspace{2.3cm}+\,A(S_{1}l_{1}s\rho,S_{2}sl_{2}\sigma,S_{1}l_{1}l_{2}\omega)|^2
\label{JPQML} \ea
\end{footnotesize}

\noindent where $P^{QM}(\rho,\sigma,\omega)$ denotes the joint
probability of getting the outcome D$_{1}(\rho)$, D$_{2}(\sigma)$,
D$_{2}(\omega)$; $A(path\,\rho,path\,\sigma,path\,\omega)$ the
corresponding probability amplitudes for the paths and outcome
triplets specified within the parentheses; and $K$ is a
normalization factor.

Substituting the amplitudes into Eq. (\ref{JPQML}) yields the
following values for the conventional joint probabilities:

\begin{footnotesize}
\ba Pr^{QM}(\rho,\sigma,\omega) =\frac{1}{8}
[1+\rho\sigma\omega\,\sin(\alpha-\beta-\gamma+\varphi_{2}-\varphi_{1})
] \label{qmjp} \ea
\end{footnotesize}

Eq. (\ref{qmjp}) yields the following correlation coefficient:

\begin{footnotesize}
\ba &&E^{QM}
=\frac{\sum_{\rho,\,\sigma,\,\omega}\rho\sigma\omega\,P^{QM}(\sigma,\omega)}
{\sum_{\rho,\,\sigma,\,\omega}P^{QM}(\rho,\sigma,\omega)}\nonumber\\
\nonumber\\
&&\hspace{9mm}=\,\sin(\alpha-\beta-\gamma+\varphi_{2}-\varphi_{1})
\label{qmcc} \ea
\end{footnotesize}

We implement now the principles and rules of Multisimultaneity
\cite{as00.1} in the context of 3-particle experiments.

We denote $T_{i1}$ the time at which the choice between reflection
and transmission occurs at device CD$_{i1}$ \cite{ophoc}. In
expressions like $(T_{i1}<T_{j1})_{i1}$ the subscript ${i1}$ after
the parenthesis denotes that all times within the parentheses are
measured in the inertial frame defined through the velocity of
choice-device CD$_{i1}$ at the instant of the choice in this
device. If it holds that $(T_{i1}<T_{j1})_{i1}$ and
$(T_{i1}<T_{k1})_{i1}$, for ($i, j, k\in\{1, 2, 3\}, i \neq j\neq
k$), the choice at CD$_{i1}$ is said to occur with ``before''
timing, and labeled $b_{i1}$. If it holds that $(T_{j1}>T_{i1}
\geq T_{k1})_{i1}$, the choice at CD$_{i1}$ is said to occur with
``after'' timing with relation to the choice at CD$_{k1}$, and
labeled $a_{i1[k1]}$. If it holds that $(T_{i1} \geq T_{j1})_{i1}$
and $(T_{i1}\geq T_{k1})_{i1}$, the choice at CD$_{i1}$ is said to
occur with ``after'' timing with relation to CD$_{j1}$ and
CD$_{k1}$, or simply with after timing, and labeled $a_{i1}$. A
before-choice at CD$_{i1}$ carrying out the value $\rho$ is
denoted $b_{i\,\rho}$, and an after-choice at CD$_{i}$ carrying
out the value $\rho$ is denoted $a_{i\,\rho}$.

The main \emph{Principles} of Multisimultaneity are the following
two:

\emph{Principle 1}: The values $b_{i1\,\rho}$ of particle $i$ do
not depend on the values the other particles may produce.

\emph{Principle 2}: The values $a_{i1\,\rho}$ involve nonlocal
causal links, and depend on the values the other particles may
produce.

Regarding \emph{Principle 2}, in after-after timings it would
obviously be absurd to assume together that $a_{i1\,\rho}$ depends
on $a_{j1\,\sigma}$, and $a_{j1\,\sigma}$ on $a_{i1\,\rho}$.
Therefore we assume that the outcomes particle $i$ produces in
after choices at CD$_{i1}$ do not depend on the outcomes the other
particles $j$ and $k$ may actually produce in after choices but on
the outcomes they would have produced if the choices at CD$_{j1}$
and CD$_{k1}$ would have been before ones.

We denote $Pr(C)$ the probability that a photon triplet belongs to
the class traveling path $C$,
$C\in\{(S_{1}sl_{1},S_{2}l_{2}s,S_{2}l_{2}l_{1}),
(S_{1}l_{1}s,S_{2}sl_{2},S_{1}l_{1}l_{2})\}$; Expressions like
$Pr(b_{i1\,\rho}, a_{j1[i1]\,\sigma}, a_{k1\,\omega})$ denote the
probabilities of getting the indicated values. $P(b_{i1\,\rho}|C)$
the conditional probability that photon $i$ leaves CD$_{i1}$ by
output port $\rho$ after a before-choice, providing the pair
travels path $C$; expressions as $P(a_{k1\,\omega}|b_{i1\,\rho},
a_{j1[i1]\sigma},)$ mean the conditional probability that photon
$i$ leaves CD$_{i1}$ by output port $\rho$ after an after-choice
providing photon $j$ would have left CD$_{j1}$ by output port
$\sigma$ in a before-choice, and photon $k$ CD$_{k1}$ by output
port $\omega$ in an after-choice with relation to CD$_{j1}$.

The rule to calculate the joint probabilities for the outcomes at
CD$_{11}$, CD$_{21}$ and CD$_{31}$, with all choices occurring
under before timing, follows straightforwardly from
\emph{Principle 1} above, and is given by the expression:

\begin{footnotesize} \ba &&Pr(b_{11\,\rho}, b_{21\,\sigma},
b_{31\,\omega})\nonumber\\
&&=\sum_{C}Pr(C)\,Pr(b_{11\,\rho}|C)\,Pr(b_{21\,\sigma}|C)\,Pr(b_{31\,\omega}|C)
 \label{PBBB} \ea
\end{footnotesize}

For the different paths it holds that:

\begin{footnotesize}
\ba &&Pr(C)=\frac
{1}{K}|A(S_{1}sl_{1},S_{2}l_{2}s,S_{2}l_{2}l_{1})|^2\nonumber\\
&&=\frac
{1}{K}|A(S_{1}l_{1}s,S_{2}sl_{2},S_{1}l_{1}l_{2})|^2=\frac {1}{2}
\label{PC} \ea
\end{footnotesize}

And for the $b_{i1\,\rho}$ values, $i\in\{1,2,3\}$,
$\rho\in\{+,-\}$, one is led to the following relations:

\begin{footnotesize}
\ba Pr(b_{i1\,\rho}|C)=
|A(s\,\rho)|^2=|A(l_{1}\,\rho)|^2=|A(l_{2}\,\rho)|^2=\frac {1}{2}
\label{Pbi1|C}\ea
\end{footnotesize}

Substituting (\ref{PC}) and (\ref{Pbi1|C}) into (\ref{PBBB})
yields:

\begin{footnotesize} \ba Pr(b_{11\,\rho}, b_{21\,\sigma},
b_{31\,\omega})=\frac {1}{8} \label{PBBBvalue} \ea
\end{footnotesize}

For each choice-device with input ports $(p, q)\in\{(l_{1}, s),
(l_{2}, s),(l_{1}, l_{2})\}$ and output ports ($+, -$), the path
amplitudes fulfill the relation:

\begin{footnotesize}
\ba &&A(p\,+)A^*(q\,+)+A(p\,-)A^*(q\,-)=0 \label{AA*} \ea
\end{footnotesize}

Relation (\ref{AA*}) implies that:

\begin{footnotesize}
\ba &&\sum_{\omega}Pr^{QM}(\rho,\sigma,\omega)\nonumber\\
&&=\sum_{\omega}|A(S_{1}sl_{1}\rho,S_{2}l_{2}s\sigma,S_{2}l_{2}l_{1}\omega)
+A(S_{1}l_{1}s\rho,S_{2}sl_{2}\sigma,S_{1}l_{1}l_{2}\omega)|^2\nonumber\\
&&=|C|^2|A(l_{1}\rho)|^2|A(s\sigma)|^2+|C|^2|A(s\rho)|^2|A(l_{2}\sigma)|^2\nonumber\\
&&=\sum_{C}Pr(C)\,Pr(b_{i1\,\rho}|C)\,Pr(b_{j1\,\sigma}|C)
=Pr(b_{i1\,\rho},b_{j1\,\sigma})
\label{PBB=PBA}\ea
\end{footnotesize}

\noindent and similar equalities for summations over $\rho$ and
$\sigma$.

Relation (\ref{PBB=PBA}) means that for the experiment we are
considering the quantum mechanical marginals can be described as
though the involved choices would be before ones, i.e. the values
$a_{i1[k1]\,\rho}$ behave as $b_{i1\,\rho}$ ones.

Consider \emph{first} the type of timing that practically result
when all choice-devices are at rest, i.e. $(b_{i1\,\rho},
a_{j1[i1]\,\sigma}, a_{k1\,\omega})$. For these timings we assume
that Multisimultaneity reproduces the predictions of Quantum
Mechanics, so that:

\begin{footnotesize}
\ba Pr(b_{i1\,\rho}, a_{j1[i1]\,\sigma}, a_{k1\,\omega})
=Pr^{QM}(\rho,\sigma,\omega) \label{PMS=PQM} \ea
\end{footnotesize}

From \emph{Principle 2} above it follows that:

\begin{footnotesize}
\ba &&Pr(b_{i1\,\rho}, a_{j1[i1]\,\sigma},
a_{k1\,\omega})\nonumber\\
&&=\sum_{C}P(C)\,P(b_{i1\,\rho}|C)\,P(a_{j1\,\sigma}|b_{i1\,\rho})\,P(a_{k1\,\omega}|b_{i1\,\rho},
a_{j1[i1]\,\sigma}) \label{PBAj1[i1]A} \hspace{7mm} \ea
\end{footnotesize}

Summing over $\omega$ in (\ref{PBAj1[i1]A}), and taking account of
(\ref{PMS=PQM}) and (\ref{PBB=PBA}) one is led to:

\begin{footnotesize}
\ba &&\sum_{\omega}Pr(b_{i1\,\rho}, a_{j1[i1]\,\sigma},
a_{k1\,\omega})=\sum_{\omega}Pr^{QM}(\rho,\sigma,\omega)\nonumber\\
&&=\sum_{C}P(C)\,P(b_{i1\,\rho}|C)\,P(A_{j1\,\sigma}|b_{i1\,\rho})\nonumber\\
&&=\sum_{C}P(C)\,P(b_{i1\,\rho}|C)\,P(b_{j1\,\sigma}|C)\label{sumPqm}
\ea
\end{footnotesize}

From (\ref{PBAj1[i1]A}) and (\ref{sumPqm}) it follows that:

\begin{footnotesize}
\ba &&Pr(a_{k1\,\omega}|b_{i1\,\rho},a_{j1[i1]\,\sigma})
=Pr(a_{k1\,\omega}|b_{i1\,\rho},b_{j1\,\sigma})\nonumber\\
&&=\frac{Pr^{QM}(\rho, \sigma, \omega)}
{\sum_{\omega}Pr^{QM}(\rho, \sigma, \omega)}\nonumber\\
&&=\frac{1}{2}[1+\rho\sigma\omega\,\sin(\alpha-\beta-\gamma+\varphi_{2}-\varphi_{1})]
\label{Pak1} \ea
\end{footnotesize}

Since by setting CD$_{j1}$ in movement one could change
instantaneously the timing ($b_{i1}, a_{j1[i1]}, a_{k1}$) into
$(b_{i1}, b_{j1}, a_{k1})$, property (\ref{PBB=PBA}) prevents that
this action can be used to produce superluminal signaling. For
experiments that don't fulfill (\ref{PBB=PBA}) Multisimultaneity
can be conveniently completed so that superluminal signaling
remains forbidden.

Consider \emph{secondly} the experiment of Fig. 1 (b) conducted
with timing $(a_{11[31]}, a_{21[31]}, b_{31})$. Since as stated
above the values $a_{i1[k1]\,\rho}$ behave as $b_{i1\,\rho}$ ones,
taking account of (\ref{PBBBvalue}) one is led to:

\begin{footnotesize}
\ba Pr(a_{11[31]\,\rho}, a_{21[31]\,\sigma}, b_{31\,\omega})
=Pr(b_{11\,\rho}, b_{21\,\sigma}, b_{31\,\omega})=\frac {1}{8}
\label{PBAj1[i1]A=PBBB}\ea
\end{footnotesize}

And (\ref{PBAj1[i1]A=PBBB}) yields the following correlation
coefficient:

\begin{footnotesize}
\ba E^{bbb}
=\frac{\sum_{\rho,\,\sigma,\,\omega}\rho\sigma\omega\,Pr(b_{11\,\rho},
b_{21\,\sigma}, b_{31\,\omega})} {\sum_{\rho,\,\sigma,\,\omega}
Pr(b_{11\,\rho}, b_{21\,\sigma}, b_{31\,\omega})} = 0
\label{BBBcc} \ea
\end{footnotesize}

Consider \emph{thirdly} the timing $(a_{11\,\rho}, a_{21\,\sigma},
b_{31\,\omega})$. Applying \emph{Principle 2} one gets:

\begin{footnotesize}
\ba &&Pr(a_{11\,\rho}, a_{21\,\sigma}, b_{31\,\omega}) \nonumber\\
&&=\sum_{C,\rho',\sigma'}Pr(C)\,Pr(b_{11\,\rho'}|C)\,Pr(b_{21\,\sigma'}|C)
Pr(b_{31\,\omega}|C)\nonumber\\
&&\hspace{1.2cm}\times\,Pr(a_{11\,\rho}|b_{21\,\sigma'},
b_{31\,\omega}) \,Pr(a_{21\,\sigma}|b_{11\,\rho'},
b_{31\,\omega})\label{PAAB}\hspace{7mm} \ea
\end{footnotesize}

Substituting (\ref{PC}), (\ref{Pbi1|C}) and (\ref{Pak1}) into
(\ref{PAAB}) gives:

\begin{footnotesize}
\ba Pr(a_{11\,\rho}, a_{21\,\sigma}, b_{31\,\omega})= \frac {1}{8}
\label{PAABvalue} \ea
\end{footnotesize}

Eq. (\ref{PAABvalue}) yields the following correlation
coefficient:

\begin{footnotesize}
\ba E^{aab}
=\frac{\sum_{\rho,\,\sigma,\,\omega}\rho\sigma\omega\,Pr(a_{11\,\rho},
a_{21\,\sigma}, b_{31\,\omega})} {\sum_{\rho,\,\sigma,\,\omega}
Pr(a_{11\,\rho}, a_{21\,\sigma}, b_{31\,\omega})} =0 \label{AABcc}
\ea
\end{footnotesize}

Consider \emph{finally} the timing $(a_{11\,\rho}, a_{21\,\sigma},
a_{31\,\omega})$. \emph{Principle 2} leads to the following rule:

\begin{footnotesize}
\ba &&Pr(a_{11\,\rho}, a_{21\,\sigma}, a_{31\,\omega}) \nonumber\\
&&=\sum_{C,\rho',\sigma'}Pr(C)\,Pr(b_{11\,\rho'}|C)
\,Pr(b_{21\,\sigma'}|C)Pr(b_{31\,\omega'}|C) \nonumber\\
&&\hspace{12mm}\times\,Pr(a_{11\,\rho}|b_{21\,\sigma'},
b_{31\,\omega'}) \,Pr(a_{21\,\sigma}|b_{11\,\rho'},
b_{31\,\omega'})\hspace{7mm}\nonumber\\ &&\hspace{12
mm}\times\,Pr(a_{31\,\sigma}|b_{11\,\rho'},
b_{21\,\sigma'})\label{PAAA} \ea
\end{footnotesize}

Substituting (\ref{PC}), (\ref{Pbi1|C}) and (\ref{Pak1}) into
(\ref{PAAA}) gives:

\begin{footnotesize}
\ba &&Pr(a_{11\,\rho}, a_{21\,\sigma}, a_{31\,\omega})\nonumber\\
&&=\frac{1}{8}[1+\rho\sigma\omega\,\sin^3(\alpha-\beta-\gamma+\varphi_{2}-\varphi_{1})]
\label{PAAAvalue} \ea
\end{footnotesize}

Eq. (\ref{PAAAvalue}) yields the following correlation
coefficient:

\begin{footnotesize}
\ba &&E^{aaa}
=\frac{\sum_{\rho,\,\sigma,\,\omega}\rho\sigma\omega\,Pr(A_{11\,\rho},
A_{21\,\sigma}, A_{31\,\omega})} {\sum_{\rho,\,\sigma,\,\omega}
Pr(A_{11\,\rho}, A_{21\,\sigma}, A_{31\,\omega})}\nonumber\\
&&=\sin^3(\alpha-\beta-\gamma+\varphi_{2}-\varphi_{1})
\label{AAAcc} \ea
\end{footnotesize}

The predictions (\ref{BBBcc}) and (\ref{AABcc}) are quantitatively
clearly testable against the prediction (\ref{qmcc}): while
Quantum Mechanics predicts a correlation coefficient $E$
oscillating between $1$ and $-1$ depending on the phase values,
Multisimultaneity predicts the constant value $E=0$.

Real tests are feasible. Before-before and after-after timings
require that the difference $\delta t$ between the optical paths
traveled by the two photons, the real distance $D$ between the two
choice-devices, and the velocity $V$ defining the inertial frame,
fulfill the relation \cite{asvs97.1}:

\begin{footnotesize}
\ba D >\frac {c^2\delta t} {V} \label{dt} \ea
\end{footnotesize}

Working within the laboratory one can achieve an alignment
ensuring that $\delta t \approx 2$ ps. Acousto-optic cells make it
possible to reach values as $V= 2.5$ km/s \cite{as00.1}. Then,
relation (\ref{dt}) yields the following lower limit for the
distance $D$ between CD$_{11}$ and CD$_{21}$:

\begin{footnotesize}
\ba D > 72\,m \label{dt2} \ea
\end{footnotesize}

To have the timing $(b_{11}, b_{21}, b_{31})$  the direction of
movement is that indicated in Fig. 1 (b), and for timing $(a_{11},
a_{21}, b_{31})$ the reverse, and moreover, in both cases,
CD$_{31}$ has to be set so that the arrival of particle 3 at
CD$_{31}$ in the laboratory frame occurs before the arrivals of
particle 1 and 2 at CD$_{11}$, respectively CD$_{21}$.

Regarding the prediction (\ref{AAAcc}), it might be difficult to
distinguish it from (\ref{qmcc}) because the corresponding plots
of the experimental data will exhibit quite similar shapes.
Nevertheless, assumed the experiment upholds (\ref{BBBcc}) and
(\ref{AABcc}), the interest of testing (\ref{AAAcc}) will of
course be that of a further confirmation of Multisimultaneity.
Timing $(a_{11}, a_{21}, a_{31})$ requires the following: to
adjust the location of CD$_{31}$ so that the arrival of particle 3
at CD$_{31}$ in the laboratory frame occurs after the arrivals of
particle 1 and 2 at CD$_{11}$, respectively CD$_{21}$; to reverse
the direction of movement indicated in Fig. 1; and to orient
CD$_{i1}$, $i\in\{1, 2\}$ till to get the sound wave propagating
towards CD$_{31}$. Assumed a value of $90^{o}$ for the angle
CD$_{11}$-CD$_{31}$-CD$_{21}$, and distances CD$_{11}$-CD$_{31}$,
CD$_{21}$-CD$_{31}$ of about 72 m, then one has a distance
CD$_{11}$-CD$_{21}$ of 102 m, and  velocity components of 1.77
km/s in the direction CD$_{11}$-CD$_{21}$, i.e. values which
fulfill the condition (\ref{dt}).

In summary, we have shown that a description imbedding nonlocal
causality in a relativistic time ordering is possible for
experiments involving more than 2 particles.

As far as one aims nonlocal causal descriptions based on
relativistic (real) timings, it is natural to assume that each
choice involves all (local and nonlocal) information that is
available within the inertial frame of the choice-device at the
instant the particle arrives. Then the basic principle of any
relativistic nonlocal description is the following one: in
experiments in which all choices take place under before timing
the nonlocal correlations should disappear. By contrast Quantum
Mechanics predicts such correlations independently of any timing.
Therefore, experiments using only before timings can be considered
a criterion allowing us to decide whether or not the nonlocal
physical reality can be imbedded into a relativistic chronology.

Prevailing of Quantum Mechanics in forthcoming Bell experiments
with moving choice-devices would support the view that the
nonlocal correlations are caused independently of any real
chronology \cite{pf}. In this sense Quantum Mechanics can be
considered a causal description which is both \emph{nonlocal} and
\emph{nontemporal}.

It is a pleasure to thank Nicolas Gisin, Valerio Scarani, Andr\'e
Stefanov, and Hugo Zbinden for very stimulating discussions, and
the Odier Foundation of Psycho-physics for financial support.

\begin{footnotesize}

\end{footnotesize}
\end{document}